\documentclass[%
reprint,
amsmath,amssymb,
aps,
]{revtex4-1}

\usepackage{graphicx}% Include figure files
\usepackage{appendix}
\usepackage{varioref}
\usepackage{cleveref}
\usepackage{subfig}
\usepackage{dcolumn}% Align table columns on decimal point
\usepackage{bm}% bold math
\usepackage[mathlines]{lineno}% Enable numbering of text and display math
\usepackage{xcolor}
\begin{document}
	
	%\preprint{APS/123-QED}
	
	\title{Introducing vortices in the continuum using direct and indirect methods}% Force line breaks with \\
	
	\author{Zahra\ Asmaee}%
	\email{{\color{blue}{zahra.asmaee@ut.ac.ir}}}
	\author{Sedigheh\ Deldar}
	\email{{\color{blue}{sdeldar@ut.ac.ir}}}
	\affiliation{%
		Department of  Physics, University of Tehran,\\
		P. O. Box 14395/547, Tehran 1439955961, Iran
	}%
\author{Motahareh\ Kiamari}%
\email{{\color{blue}{mkiamari@ipm.ir}}}
\affiliation{%
	School of Particles and Accelerators, IPM,\\
	P. O. Box 19395-5531, Tehran, Iran
}%
	\begin{abstract}
		Inspired by direct and indirect maximal center gauge methods which confirm the existence of vortices in lattice calculations and by using the connection formalism, we show that under some appropriate gauge transformations vortices and chains appear in the QCD vacuum of the 
	continuum limit. 
	In the direct method, by applying center gauge transformation and \textquotedblleft center projection,\textquotedblright QCD is reduced to a gauge theory including 
	vortices, which corresponds to the non-trivial first homotopy group $\Pi_1\left( \text{SO}(3)\right) =Z_2.$  On the other hand, 
	using the indirect method, in addition to the center gauge transformation and \textquotedblleft center projection,\textquotedblright an initial step called 
	Abelian gauge 
	transformation and then Abelian projection are applied. Therefore, instead of single vortices, chains that contain monopoles and vortices 
	appear in the theory.

		\vspace*{0.2cm}
		\begin{description}
			\item[PACS numbers]
			12.38.Aw, 12.38.Lg, 12.40.-y.
		\end{description}
	\end{abstract}
	
	\pacs{Valid PACS appear here}% PACS, the Physics and Astronomy
	\maketitle
	
	%\tableofcontents
	
	\section{\label{sec:level1}Introduction}

Quantum chromodynamics is the non-Abelian gauge theory of the strong interaction which describes the hadrons in terms of quarks and gluons. There are many books that discuss QCD; for instance, see Refs. \cite{QCD1986, QCD2010}.
However, quarks have not been observed as isolated particles in the real world. Only hadrons (mesons and baryons) are observed as some color singlet 
combinations.
This experimental fact reflects the confinement mechanism as one of the most controversial unsolved issues in particle physics in the low-energy regime or large distances \cite{confinement2004,conf2011}.
During the past decades, many ideas have been proposed to approach this problem. There are many articles about this subject; for instance, see Refs. \cite{aconfinement1979, a2confinement1999, a3confinement1999,aconfinement1980, aconfinement1981, aconfinement2005, aconfinement2008}.

The area law of the Wilson loop average is a well-known gauge-invariant criterion in studying quark confinement. It leads to a linear potential between a pair of static quark-antiquark.
To study the linear part of the confinement potential, the quenched approximation is used where the dynamical quarks are removed for the infrared regime \cite{conf2011}.
In fact, one can obtain some collective modes from gluons \cite{Ichie} which are associated with some topological degrees of freedom of the QCD vacuum, and as a result it is assumed that the QCD vacuum is filled with the topological objects obtained from these collective modes.  Magnetic monopoles and center vortices are among the main candidates for describing confinement and each has its own fans.

For the non-perturbative description, people use lattice QCD simulations and phenomenological models to look for the confinement and topological objects. 
The results of the phenomenological models must be in agreement with the results of the lattice QCD, though. In fact, lattice QCD can be served as a laboratory for confirming the correctness or incorrectness of the phenomenological models.

In the absence of matter fields, some various mechanisms of confinement have been suggested to extract the topological degrees of freedom of pure Yang-Mills theory. 
One of those mechanisms is the picture of the dual superconductor and appearance of Abelian monopoles.
It was proposed by Nambu \cite{Nambu}, Mandelstam \cite{Mandelstam}, 't Hooft \cite{'t Hooft} and  
Polyakov \cite{Polyakov} in the 1970s. The idea is that the QCD vacuum can behave like a dual superconductor and it is filled with magnetic monopoles. 
Just as the Meissner effect leads to the condensation of the Cooper pairs as electrically charged objects in an ordinary superconductor, the magnetic monopoles are condensed in a dual superconductor and squeeze the chromoelectric flux between the quark-antiquark pair inside a tube. Therefore, confinement of electric fields is obtained as a result of the condensation of magnetic monopoles in this picture \cite{Ichie, Ripka, kondo}.

The second possible mechanism is given by the center vortex model \cite{DFG97,FGO,DFG98,Greensite,DFGO97,DFGO,AGG}.
Historically, vortex-like structures were introduced in superconductors in 1959. Even though they were not observed at that time, they were recognized a few years later by Abrikosov \cite{Abrikosov}.
It was proposed in various forms by 't Hooft \cite{t Hooft78, t Hooft80,t Hooft81,t Hooft82}, Nielsen and Olesen \cite{NO}, Ambjorn and Olesen \cite{AO}, Mack and Petkova \cite{Mack,MP}, and 
Cornwall \cite{Cornwall} in the late 1970s with a field theoretical approach. The idea is that the QCD vacuum is filled with closed magnetic vortices, and it 
is assumed that the vortices are condensed in the QCD vacuum. If a Wilson loop is linked to a vortex in an SU($N$) gauge group, the Wilson loop obtains 
a phase difference equal to $e^{i2\pi n/N}$ ($n=0$ to $N-1$) corresponding to the type of the vortex. As a result,  
some disorders are created in the lattice which eventually lead to an area law fall-off and confinement. 

Vortices are defined by the center of the SU($N$) gauge group and there exist $(N-1)$ distinct vortices, which are called non-Abelian Z$_{N}$ vortices. 
The simplest vortices are defined by the Z$_2$ gauge group and they have the topology of tubes (in three Euclidean dimensions) or surfaces of finite thickness (in four dimensions) carrying some well-defined magnetic fluxes \cite{aconfinement1979,AO,Mack,MP,Cornwall,t Hooft78,t Hooft80,NO}.

Lattice calculations show that Z$_{N}$ vortices produce full string tensions as the Yang-Mills vacuum does. This is 
an encouraging motivation to study confinement via center vortices. If the center vortices are removed from the lattice, the string tension also disappears 
\cite{DFGO97,DFG97,DFG98,LTER,ELRT,GLR}.

The vortex condensation picture relies upon center gauge fixing and center projection. After performing center projection in lattice QCD, the full QCD with 
SU($N$) gauge symmetry is reduced to a gauge theory with a Z($N$) gauge symmetry. These vortices are called projection vortices (or p-vortices). 
Unlike monopoles, the modified models of vortices like thick center vortices can qualitatively explain the Casimir scaling dependence for all representations \cite{FGO}.

To study the confinement problem by center vortices, one first has to discuss the existence of vortices in the continuum limit.
The most common methods of identifying vortices in the lattice simulation are direct maximal center gauge (DMCG) \cite{DFG98} and indirect maximal center gauge (IMCG)  \cite{DFG97}.
Inspired by these two methods of identifying vortices in lattice calculations and by the help of the connection formalism \cite{Ichie}, we discuss the appearance of vortices in the continuum limit of QCD.

We review DMCG and IMCG methods in lattice QCD in Sec. \ref{sec:level2}.
In Sec. \ref{sec:level3}, motivated by the methods proposed in lattice calculations, we introduce the vortices in the continuum by direct method for SU($N$) gauge group. As an example, by applying an appropriate gauge transformation in the SU($2$) gauge group and using the results of 
Sec. \ref{sec:level3}, we show in Sec. \ref{sec:level4} that under the center gauge transformation the vortex and anti-vortex can appear in the theory. 
Then, we remove the term that represents 
the anti-vortex. The theory has an SO($3$) symmetry containing the vortex, which corresponds to the non-trivial first homotopy group of $\Pi_1\left( \text{SO}(3)\right) =Z_2.$ 
Removal of the contribution of the anti-vortex is called 
\textquotedblleft center projection\textquotedblright\ in our paper.  
In Sec. \ref{sec:level5}, we introduce thin vortices in the continuum by the indirect method for SU($N$) gauge group.
Sec. \ref{sec:level6} is brought in two subsections.
In Sec. \ref{sub 6.1}, applying an Abelian gauge transformation for SU($2$) gauge theory, we show that the QCD vacuum is 
filled with monopoles and anti-monopoles. It is shown that after Abelian projection the monopole 
appears in the vacuum and the gauge group symmetry is reduced from SU($2$) to U($1$) and we have a monopole vacuum.
Then, in Sec. \ref{sub 6.2} we show that under a center gauge transformation on the monopole vacuum the vortex and anti-vortex 
appear in the gauge theory along with the monopole. After applying a \textquotedblleft center projection\textquotedblright\ we have a gauge theory 
that contains chains including monopoles and vortices.

\section{\label{sec:level2}DMCG and IMCG in lattice QCD}

There are some methods to identify vortices in lattice calculations and by using appropriate gauge transformations, which are in agreement with the vortex condensation picture \cite{DFG97,FGO,DFG98,DFGO97,DFGO}.

In lattice QCD, the action is expressed in terms of link variables on which the gluon fields are defined. The idea is that under an appropriate gauge transformation the link variables $U_\mu(x)$ get as close as possible to the center gauge group;  $\text{center}\left( \text{SU}(N)\right) =Z_N$. Then, after a projection, 
a smaller set of degrees of freedom remains. This job is usually done via two methods in lattice QCD calculation. In the following, we briefly review both methods.
\subsection{Direct maximal center gauge method }\label{sub 2.1}
This method was proposed by Del Debbio \textit{et al.} \cite{DFG98}, who tried to maximize the following quantity by determining the gauge transformation $G(x)\in \text{SU}(N)$:
\begin{equation}
	R[U]=\underset{G}{\text{max}} \sum_{x,\mu} \lvert \text{Tr}\  U^{G}_\mu(x)\lvert^2.
	\label{1}
\end{equation}
$U^{G}_\mu(x)=G(x)U_\mu(x) G^{\dagger}(x+\hat{\mu})$ shows the gauge transformation of the link variables $U_{\mu}(x)$ and as a result of the above maximization, 
$U^{G}_\mu$ becomes as close as possible to the center elements.
$\hat{\mu}$ is a unit vector along the $\mu$ direction. Then, by performing the center projection, one replaces the transformed link variable $U^{G}_\mu(x)$ 
by the closest associated center elements of the group 
Z$_N$. As an example, the center projection is defined for the SU($2$) gauge group as
\begin{equation}
	U^{G}_\mu(x)\rightarrow Z(2)=\text{sign}\left[ \text{Tr}\  U^{G}_\mu(x) \right] \textbf{1} =\left\lbrace +1,-1\right\rbrace \times \textbf{1},
	\label{2}
\end{equation}
where $\textbf{1}$ represents a $2\times2$ unit matrix. P-vortices identified by the DMCG method, are related to the non-perturbative degrees of freedom; see Ref. \cite{conf2011} for more details.

\subsection{Indirect maximal center gauge method}\label{sub 2.2}

The indirect maximal center gauge method was originally examined in lattice QCD for the SU($2$) gauge group \cite{DFG97}. 
In general, for an SU($N$) gauge group, the procedure is as the follows. For the first step, a non-Abelian gauge configuration is fixed under Abelian gauge 
fixing, and after Abelian projection, the SU($N$) gauge symmetry is reduced to $\left[\text{U}(1) \right] ^{N-1 }$. In the second 
step, the remaining $\left[\text{U}(1) \right] ^{N-1 }$ configuration is fixed under center gauge fixing such that the transformed gauge 
fields become as close as possible to the center elements. Finally, by performing a center projection, one gets the center elements. For example, for the SU($2$) gauge group,
\begin{equation}
	Z(2)=\sum_{x,\mu}\text{sign}\left[cos\theta(x,\mu)\right] \textbf{1} =\left\lbrace +1,-1\right\rbrace\times \textbf{1},
	\label{3}
\end{equation}
where $\theta(x,\mu)$ parametrizes the links.
In general, in both methods, identification of the vortices is done by using gauge fixing and then projection.

\section{\label{sec:level3} direct method of introducing vortices in the Continuum}

In this section, inspired by DMCG  method in lattice QCD which confirms the existence of vortices
in the infrared regime, we show that vortices appear in the theory by a singular
gauge transformation. In this procedure, we use the connection formalism, which is applied for the singular gauge transformation.
We would like to mention that we do not find a continuum
formula for maximal center gauge transformation, Eq.(\ref{1}), which has already been done in Ref. \cite{ER2017}. Instead, motivated by the fact
that vortices exist in the infrared part of the theory, as shown by the lattice
calculations, we use the connection formalism to make explicit the vortices, somehow similar to the procedure done in Ref. \cite{Ichie} for monopoles.

\subsection{Link variable and transformed gluon field}\label{sub 3.1}
In lattice gauge theory, color confinement can be studied by quenched approximation where only gluon fields exist in the theory. Gluons are defined on link variables as follows:
\begin{equation}
	U_\mu(x)=e^{iagA_{\mu}(x)} \in \text{SU}(N).
	\label{4}
\end{equation}
Under a gauge transformation $G(x)\in \text{SU}(N)$, the link variables are transformed as
\begin{equation}
	U_\mu(x)\rightarrow U^G_\mu(x)=G(x)U_\mu(x) G^{\dagger}(x+\hat{\mu}).
	\label{5}
\end{equation}
Using Eq.\eqref{4} in the above equation,
\begin{equation}
	U^G_\mu(x)=G(x)e^{iagA_{\mu}(x)}G^{\dagger}(x+\hat{\mu}).
	\label{6}
\end{equation}
Since the lattice spacing $a$ is small enough, we can use the exponential expansion of $e^{iagA_{\mu}(x)},$ and using the Taylor expansion for $G^{\dagger}(x+\hat{\mu})$,
\begin{equation}\begin{split}		
			U^G_\mu&=1+iag\left[G(x)A_\mu G^{\dagger}(x)-\dfrac{i}{g} G(x)\partial_\mu G^{\dagger}(x)\right] +\mathcal{O}(a^2)
			\\
			&=e^{iagA^G_{\mu}}.
			\label{7}
\end{split}\end{equation}
Thus, in the continuum limit where $a\rightarrow 0$, the gluon field is transformed as
\begin{equation}
	A^G_\mu(x)=G(x)A_\mu(x) G^{\dagger}(x)-\dfrac{i}{g} G(x)\partial_\mu G^{\dagger}(x),
	\label{8}
\end{equation}
where $A^G_\mu(x)\in \text{SU}(N)$. In terms of group generators,
\begin{equation}
	\vec{A}^G_\mu.\vec{T}=G(x)\left( A^c_{\mu}T^c\right)  G^{\dagger}(x)-\dfrac{i}{g} G(x)\partial_\mu G^{\dagger}(x),
	\label{9}
\end{equation}
$T^c$ are generators of the SU($N$) group, and $c$ is the color index.

Since we are interested in observing topological defects from the gluon fields, we have to use an appropriate gauge transformation. Thin vortices appear 
as topological defects after center gauge transformations and some subsequent efforts.

Equation \eqref{9} can be used to study the vortices if $G(x)\equiv N(x)$ is defined as a center gauge transformation,
\begin{equation}
	\vec{A}^N_\mu.\vec{T}=N(x)\left( A^c_{\mu}T^c\right)  N^{\dagger}(x)-\dfrac{i}{g} N(x)\partial_\mu N^{\dagger}(x).
	\label{10}
\end{equation}
To study the contribution of thin vortices in the continuum, we first recall that in lattice QCD calculations when the Wilson loop links to the vortex
it receives a phase difference equal to $e^{i2\pi n/N}$ associated with the non-trivial center element contribution Z$(k)$ \cite{kondo},
\begin{equation}
	W(C)\rightarrow e^{i2\pi n/N}W(C),\qquad(n=1,2,...,N-1).
	\label{11}
\end{equation}
\begin{figure}[ht]
	\begin{center}
		\centering
		\subfloat[]{\includegraphics[height=2.7cm, width=3.7cm]{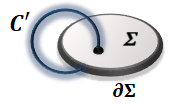}}
		\quad
		\subfloat[]{\includegraphics[height=2.7cm, width=3.7cm]{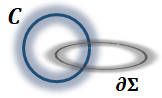}}
		\caption{Linking between the Wilson loop $C$ and a vortex with hypersurface $\Sigma$. The boundary of the hypersurface
                         is shown by $\partial\Sigma$, indicating a thin vortex.}
		\label{fig1}
	\end{center}
\end{figure}

Under a center gauge transformation $N(x)$, a Wilson line should be transformed as
\begin{equation}
	\begin{split}
		W(C^{\prime})\rightarrow W^N(C^{\prime})&=N(x-\epsilon)W(C^{\prime})N^{\dagger}(x+\epsilon)\\
		&=N(x-\epsilon)N^{\dagger}(x+\epsilon)+\mathcal{O}(\epsilon)\\
		&\equiv Z(k)+\mathcal{O}(\epsilon),
		\label{12}
\end{split}\end{equation}
$W(C^{\prime})=1+\mathcal{O}(\epsilon)$. $C^{\prime}$ indicates an open circle from $x-\epsilon$ to $x+\epsilon$, where $x$ indicates the location 
of the intersection of $C^{\prime}$ and the hypersurface $\Sigma$. $\epsilon$ is an infinitesimal quantity
so that in the limit where $\epsilon \rightarrow 0$, $C^{\prime}=C$.

We use Eq.\eqref{12} to obtain the appropriate gauge transformation $N(x)$, which gives the non-trivial center elements indicating the existence of vortices
in the last line of Eq.\eqref{12}.  

From Fig. \ref{fig1}(a), an ideal vortex is defined on $(D-1)$-dimensional hypersurface $\Sigma$, while the thin vortex is defined on $(D-2)$-dimensional boundary $S=\partial\Sigma$ \cite{kondo, ER2017}.
 Piercing the hypersurface by the Wilson loop results in a discontinuity Z$(k)$. 
The center vortex in $D=2, 3,$ and $4$ is defined as a string, surface, and volume, respectively.

The relation between an ideal vortex and a thin vortex is as follows \cite{ER2017, kondo}:
\begin{equation}
	\text{ideal\ vortex}=-\dfrac{i}{g}N(x)\partial_\mu N^{\dagger}(x)-\text{thin\ vortex}.
	\label{13}
\end{equation}
In fact, intersecting the hypersurface $\Sigma$ of an ideal vortex with a Wilson loop $C$ gives a phase to the Wilson loop proportional to a center group 
element. The boundary $\partial \Sigma$ indicates the location of the thin vortex, which is gauge equivalent to the ideal vortex. 
Thus, the ideal vortex field is not unique and can be gauge transformed to a thin vortex field which has the support only on the boundary $\partial\Sigma$ \cite{kondo}.

Replacing $\left( -\dfrac{i}{g}N(x)\partial_\mu N^{\dagger}(x)\right)$ from Eq.\eqref{13} in Eq.\eqref{10},
\begin{equation}
	\vec{A}^N_\mu.\vec{T}=N(x)\left( A^c_{\mu}T^c\right)  N^{\dagger}(x)+\text{ideal\ vortex}+\text{thin\ vortex}.
	\label{14}
\end{equation} 
On the other hand, in analogy to the lattice calculation where the thin vortex links to the Wilson loop [Fig.\ref{fig1}(b)], one can define a
gauge field in the coset space by removing the ideal vortex \cite{ER2017} so that $\vec{A}^N_\mu.\vec{T}\rightarrow \vec{A}^{\prime N}_\mu.\vec{T}$;
	\begin{equation}
			\vec{A}^{\prime N}_\mu.\vec{T}=N(x)\left( A^c_{\mu}T^c\right)  N^{\dagger}(x)+\text{thin\ vortex}.
			\label{15}
	\end{equation}
We recall that $\vec{A}^{\prime N}_\mu.\vec{T}$ is still singular on $\partial \Sigma$. The gauge field configuration $\vec{A}^{\prime N}_\mu.\vec{T}$ 
induces the same behavior for arbitrary Wilson loop as the configuration $\vec{A}^N_\mu.\vec{T}$ does.
In other words, for $x\notin$ hypersurface, we only see the boundary of the vortex, called the thin vortex field. In fact, by this choice of $x$, we 
have removed the hypersurface $\Sigma$ from the space-time. So, the contribution of the ideal vortex defined on the hypersurface vortex would be zero;
\begin{equation}
	\text{thin\ vortex}=-\dfrac{i}{g} N(x)\partial_\mu N^{\dagger}(x),\quad x\notin \text{hypersurface}.
	\label{16}
\end{equation}

\subsection{Field strength tensor and connection formalism}\label{sub 3.2}

In this subsection, we discuss the connection formalism, which has already been used in some references, for instance, Refs. \cite{Ichie, Suzuki}.
In fact, we generalize the connection formalism, previously applied to the Abelian gauge transformation, to the center gauge transformation. 

The Yang-Mills Lagrangian has an SU($N$) symmetry and is given by
\begin{equation}
	\pounds_{\text{YM}}=-\dfrac{1}{2} Tr(F_{\mu\nu}F^{\mu\nu}),
	\label{17}
\end{equation}
where the SU($N$) non-Abelian field strength tensor called $F_{\mu\nu}=\vec{F}_{\mu\nu}.\vec{T}=F^c_{\mu\nu}T^c$ is defined by 
$F_{\mu\nu}=\partial_\mu A_\nu-\partial_\nu A_\mu+ig\left[A_\mu, A_\nu\right] $ 
and for a regular system can be written as
\begin{equation}
	F_{\mu\nu}=\dfrac{1}{ig}\left[ \hat{D}_\mu,\hat{D}_\nu\right], \quad \hat{D}_\mu=\hat{\partial}_\mu+igA_\mu,
	\label{18}
\end{equation}
where $\hat{D}_\mu$ is the covariant-derivative operator.

But topological defects appear as a result of singular gauge transformation. To observe these defects explicitly, we rewrite the Yang-Mills gauge theory 
in terms of the covariant-derivative operator $\hat{D}_\mu$ and the ordinary derivative operator $\hat{\partial}_\mu$,

\begin{equation}
	\begin{split}
		\dfrac{1}{ig}\left[ \hat{D}_\mu,\hat{D}_\nu\right] &=\dfrac{1}{ig}\left[ \hat{\partial}_\mu+igA_\mu, \hat{\partial}_\nu+igA_\nu\right] 
		\\
		&=\dfrac{1}{ig}\left[ \hat{\partial}_\mu, \hat{\partial}_\nu\right]+ \left[ \hat{\partial}_\mu, A_\nu\right] +\left[ A_\mu,\hat{\partial}_\nu\right]
		\\
		&+ig\left[A_\mu,A_\nu \right].
		\label{19}
	\end{split}
\end{equation}
Using the $\left[\hat{\partial}_\mu,f \right]=\partial_\mu f $,
\begin{equation}
	\dfrac{1}{ig}\left[ \hat{D}_\mu,\hat{D}_\nu\right]=\dfrac{1}{ig}\left[ \hat{\partial}_\mu, \hat{\partial}_\nu\right]+ \partial_\mu A_\nu-\partial_\nu A_\mu +ig\left[A_\mu, A_\nu \right].
	\label{20}
\end{equation}
For regular systems, the first term on the right-hand side of Eq.\eqref{20} is zero, so we have Eq.\eqref{18}. But this term is not zero for singular systems. Therefore,
\begin{equation}
	F_{\mu\nu}=\dfrac{1}{ig}\left[ \hat{D}_\mu,\hat{D}_\nu\right]-\dfrac{1}{ig}\left[ \hat{\partial}_\mu, \hat{\partial}_\nu\right],
	\label{21}
\end{equation}
where $F_{\mu\nu}$ is the SU($N$) non-Abelian field strength tensor, and Eq.\eqref{21} is applied when the singularity exists in the system. As a result of singular 
gauge transformation, topological defects like monopoles and vortices appear in the theory.

We study the behavior of the non-Abelian field strength tensor under singular gauge transformations. 

In general, if $G(x)\in \text{SU}(N)$ represents a regular gauge transformation, the field strength tensor is transformed as $F^G_{\mu\nu}=G(x)F_{\mu\nu}G^{\dagger}(x)$. 
Therefore, $F_{\mu\nu}$ is
\begin{equation}\begin{split}
		F^G_{\mu\nu}&=G(x)\left\lbrace \partial_\mu A_\nu-\partial_\nu A_\mu+ig\left[A_\mu, A_\nu\right]\right\rbrace G^{\dagger}(x)
		\\
		&= \left( \partial_\mu A^{ G}_\nu-\partial_\nu A^{G}_\mu\right)  +ig\left[A^{ G}_\mu,A^{G}_\nu \right].
		\label{22}
\end{split}\end{equation}
On the other hand, for a singular system where $F_{\mu\nu}$ is defined by Eq.\eqref{21}, one gets
\begin{equation}
	\begin{split}
		F^G_{\mu\nu}&=\dfrac{1}{ig}G(x)\left[ \hat{D}_\mu,\hat{D}_\nu\right]G^{\dagger}(x)-\dfrac{1}{ig}G(x)\left[ \hat{\partial}_\mu, \hat{\partial}_\nu\right]G^{\dagger}(x)
		\\
		&=\dfrac{1}{ig}\left[ \hat{D}^G_\mu,\hat{D}^G_\nu\right]-\dfrac{1}{ig}G(x)\left[ \hat{\partial}_\mu, \hat{\partial}_\nu\right]G^{\dagger}(x),
		\label{23}
	\end{split}
\end{equation}
where $ \hat{D}^G_\mu=G(x) \hat{D}_\mu G^{\dagger}(x)$. The first term of Eq.\eqref{23} can be written by the help of Eq.\eqref{20} for a gauge transformed field,
\begin{equation}
	\dfrac{1}{ig}\left[ \hat{D}^G_\mu,\hat{D}^G_\nu\right]=\dfrac{1}{ig}\left[ \hat{\partial}_\mu, \hat{\partial}_\nu\right]+\partial_\mu A^G_\nu-\partial_\nu A^G_\mu +ig\left[A^G_\mu, A^G_\nu \right].
	\label{24}
\end{equation}
Replacing Eq.\eqref{24} in Eq.\eqref{23}, 
\begin{equation}
	F^G_{\mu\nu}= \left( \partial_\mu A^{ G}_\nu-\partial_\nu A^{G}_\mu\right)  +ig\left[A^{ G}_\mu,A^{G}_\nu \right]+\dfrac{i}{g}G\left[ {\partial}_\mu, {\partial}_\nu\right]G^{\dagger}.
	\label{25}
\end{equation}
This is a noticeable result. The last term of Eq.\eqref{25} shows the difference between this equation and Eq.\eqref{22}.

The advantage of using the connection formalism technique is that the gauge theory will remain gauge invariant after the singular gauge transformation. 
Equation \eqref{25} is valid for both the Abelian and center gauge transformations.
It has already been discussed for the Abelian gauge transformation \cite{Ichie, Suzuki} and we intend to use it for the center gauge transformation, as well.

If one uses Eq.\eqref{25} without applying any projection, a full QCD will be obtained at the end. We discuss how we perform
\textquotedblleft center projection\textquotedblright\ in Sec. \ref{sec:level4}.
	
\section{\label{sec:level4} direct method for introducing vortices in SU($2$) gauge group}

The formation of center vortices in the QCD vacuum relies upon two steps: center gauge transformation and \textquotedblleft center projection\textquotedblright. Using the results of Sec. \ref{sec:level3}, we discuss these two steps for the SU($2$) gauge group.

\paragraph*{\textbf{Step 1: Center gauge transformation}}

In general, a $2\times2$ gauge transformation $G(x)\in \text{SU}(2)$ is written in terms of three Euler angles $\alpha$,$\beta$,$\gamma$,
\begin{equation}
	\begin{split}
		&G(x)=e^{i\gamma(x)T^3}e^{i\beta(x)T^2}e^{i\alpha(x)T^3}
		\\
		&=\left( \begin{matrix}
			e^{\dfrac{i}{2}\left[\gamma(x)+\alpha(x) \right]} cos\dfrac{\beta(x)}{2}& e^{\dfrac{i}{2}\left[\gamma(x)-\alpha(x) \right]} sin\dfrac{\beta(x)}{2}\\ -e^{-\dfrac{i}{2}\left[\gamma(x)-\alpha(x) \right]} sin\dfrac{\beta(x)}{2} & e^{-\dfrac{i}{2}\left[\gamma(x)+\alpha(x) \right]} cos\dfrac{\beta(x)}{2}
		\end{matrix}\right)
		\\
		&\alpha(x)\in\left[ 0,2\pi\right),\ \beta(x)\in \left[ 0,\pi\right],\ \gamma(x)\in\left[ 0,2\pi\right),\ T^c=\dfrac{\sigma^c}{2},
		\label{26}
	\end{split}
\end{equation}
where $T^c$'s are generators of the SU($2$) group and $\sigma^c$'s are Pauli matrices.
The center gauge transformation $G(x)\equiv N(x)\in \text{SU}(2)$ is continuous everywhere except at the hypersurface of the vortex.
Therefore, the Euler angles are selected in a way that the constraint of Eq.\eqref{12} is satisfied. 
There are different choices for the angles. One can choose $\alpha=\gamma=\dfrac{\varphi}{2}$ and $\beta=0$,
\begin{equation}
	N=\left( \begin{matrix}
		e^{i\dfrac{\varphi}{2}}&0\\ 0& e^{-i\dfrac{\varphi}{2}}
	\end{matrix}\right), \quad \varphi\in\left[ 0,2\pi\right).
	\label{27}
\end{equation}
It can be shown that
\begin{equation}
	N(\varphi=\epsilon)N^{\dagger}(\varphi=2\pi-\epsilon)=-\textbf{1}_{2\times 2}\in Z(2),\ \epsilon\rightarrow0,
	\label{28}
\end{equation}
where $\left( -\textbf{1}_{2\times 2}\right)$ represents the non-trivial contribution of the Z($2$) gauge group.
Thus, the contribution of an ideal vortex is observed at $\varphi=0$. On the other hand, outside the hypersurface, the contribution 
of the thin vortex is represented by a pure gauge shown in  Eq.\eqref{16},
\begin{equation}
\text{thin\ vortex}\equiv  \vec{V}_\mu.\vec{T}=-\dfrac{i}{g}N\partial_\mu N^{\dagger}=-\dfrac{1}{g}\partial_\mu\varphi T^3.
	\label{29}
\end{equation}
The spatial component of thin vortex is
\begin{equation}
	\vec{V}_{\varphi}.\vec{T}=-\dfrac{g^{-1}}{\rho}T^3,\quad  \vec{V}_{\rho}.\vec{T}=0.
	\label{e30}
\end{equation}
Equation \eqref{e30} represents the gauge field associated with the thin vortex in cylindrical coordinates. The thin vortex is observed at $\rho=0$ \cite{ER2017} in the third direction of color space.
Under a center gauge transformation, the gluon field is defined by Eq.\eqref{15},
\begin{equation}
	\vec{A}^{\prime N}_\mu.\vec{T}=N\left( \sum^{3}_{c=1} A^c_\mu T^c\right) N^{\dagger}+\text{thin\ vortex},
	\label{30}
\end{equation}
where the first term on the right-hand side is regular and the second term indicates a topological defect.
Replacing Eqs.\eqref{27} and \eqref{29} in Eq.\eqref{30}, one obtains
\begin{equation}\begin{split}
			\vec{A}^{\prime N}_\mu.\vec{T}&=\left[ A^1_\mu cos\varphi+A^2_\mu sin\varphi\right]T^1\\&+\left[ -A^1_\mu sin\varphi +A^2_\mu cos\varphi\right]T^2\\&+\left[ A^3_\mu-\dfrac{1}{g}\partial_\mu\varphi\right]T^3
			\label{31}
\end{split}\end{equation}

The magnetic vortex flux $\Phi^{\text{flux}}$ is
\begin{equation}\begin{split}
		\Phi^{\text{flux}}&=\int dx^{\mu}\left( \vec{V}_{\mu}.\vec{T}\right) =-\dfrac{1}{2g}\int^{2\pi}_0  d\varphi \left( \begin{matrix}
			1& 0\\ 0& -1
		\end{matrix}\right)
	\\& =-\dfrac{2\pi}{g}T^3.
		\label{32}
\end{split}\end{equation}

The total contribution of the magnetic flux is in the third direction in color space.

Using Eq.\eqref{25} of Sec. \ref{sub 3.2} for the transformed field strength,
\begin{widetext}\begin{equation}
		\vec{F}^N_{\mu\nu}.\vec{T}= \left( \partial_\mu \left( \vec{A}^{\prime N}_\nu.\vec{T}\right) -\partial_\nu \left( \vec{A}^{\prime N}_\mu.\vec{T}\right) \right)+ig\left[\vec{A}^{\prime N}_\mu.\vec{T},\vec{A}^{\prime N}_\nu.\vec{T} \right]+\dfrac{i}{g}N\left[ {\partial}_\mu, {\partial}_\nu\right]N^{\dagger}.
		\label{33}
\end{equation}\end{widetext}
With the help of Eq.\eqref{31}, we rewrite the first term of Eq.\eqref{33} which is linear in terms of $\vec{A}^{\prime N}_{\mu}.\vec{T}$,
\begin{widetext}\begin{equation}\begin{split}
			F&^{\text{linear}}_{\mu\nu}=	\vec{F}^{\text{linear}}_{\mu\nu}.\vec{T}\equiv  \partial_\mu \left( \vec{A}^{\prime N}_\nu.\vec{T}\right) -\partial_\nu \left( \vec{A}^{\prime N}_\mu.\vec{T}\right) 
			\\
			&=\left[  \left( \partial_\mu A^1_\nu-\partial_\nu A^1_\mu\right) cos\varphi +\left( \partial_\mu A^2_\nu-\partial_\nu A^2_\mu\right) sin\varphi
			\right] T^1+	\left[ - \left( \partial_\mu A^1_\nu-\partial_\nu A^1_\mu\right) sin\varphi +\left( \partial_\mu A^2_\nu-\partial_\nu A^2_\mu\right) cos\varphi
			\right]  T^2
			\\
			&+\left[ \partial_\mu A^3_\nu-\partial_\nu A^3_\mu\right] T^3+\left[-g\left( A^1_\nu \dfrac{1}{g}\partial_\mu\varphi -A^1_\mu\dfrac{1}{g} \partial_\nu\varphi\right) sin\varphi+g \left( A^2_\nu\dfrac{1}{g} \partial_\mu\varphi -A^2_\mu\dfrac{1}{g} \partial_\nu\varphi\right) cos\varphi \right] T^1
			\\
			&+\left[ -g\left( A^1_\nu \dfrac{1}{g}\partial_\mu\varphi -A^1_\mu\dfrac{1}{g} \partial_\nu\varphi\right) cos\varphi-g \left( A^2_\nu \dfrac{1}{g}\partial_\mu\varphi -A^2_\mu \dfrac{1}{g}\partial_\nu\varphi\right) sin\varphi\right] T^2+\left( -\dfrac{1}{g}\left[ \partial_\mu, \partial_\nu\right] \varphi \right) T^3.
			\label{34}
\end{split}\end{equation}\end{widetext}
The first three sets of brackets of Eq.\eqref{34} are regular, and the fourth and the fifth sets of brackets indicate some kind of interactions between thin vortex and the off-diagonal gluon fields. 
$\left( -\dfrac{1}{g} \left[\partial_\mu,\partial_\nu \right]\varphi \right)T^3$  represents the field strength of a thin vortex field carrying a magnetic flux, which is equal to $\Phi^{\text{flux}}=-\dfrac{2\pi}{g}T^3$.
The second term of Eq.\eqref{33} can be written with the help of Eq.\eqref{31},
\begin{widetext}\begin{equation}\begin{split}
			F&^{\text{bilinear}}_{\mu\nu}=\vec{F}^{\text{bilinear}}_{\mu\nu}.\vec{T}\equiv ig\left[\vec{A}^{\prime  N}_\mu.\vec{T},\vec{A}^{\prime N}_\nu.\vec{T} \right]
			\\
			&=\left[ g\left( A^1_\mu A^3_\nu-A^1_\nu A^3_\mu\right) sin\varphi-g\left( A^2_\mu A^3_\nu-A^2_\nu A^3_\mu\right) cos\varphi\right] T^1+\left[ g\left( A^1_\mu A^3_\nu-A^1_\nu A^3_\mu\right) cos\varphi+g\left( A^2_\mu A^3_\nu-A^2_\nu A^3_\mu\right) sin\varphi\right] T^2
			\\
			&+ g\left[ A^2_\mu A^1_\nu-A^1_\mu A^2_\nu \right] T^3-\left[-g\left( A^1_\nu \dfrac{1}{g}\partial_\mu\varphi -A^1_\mu\dfrac{1}{g} \partial_\nu\varphi\right) sin\varphi+g \left( A^2_\nu\dfrac{1}{g} \partial_\mu\varphi -A^2_\mu\dfrac{1}{g} \partial_\nu\varphi\right) cos\varphi \right] T^1
			\\
			&-\left[ -g\left( A^1_\nu \dfrac{1}{g}\partial_\mu\varphi -A^1_\mu\dfrac{1}{g} \partial_\nu\varphi\right) cos\varphi-g \left( A^2_\nu \dfrac{1}{g}\partial_\mu\varphi -A^2_\mu \dfrac{1}{g}\partial_\nu\varphi\right) sin\varphi\right] T^2.
			\label{35}
\end{split}\end{equation}\end{widetext}
The first three brackets of Eq.\eqref{35} represent interactions between gluon fields and are regular. The fourth and the fifth brackets indicate interactions between the thin vortex and off-diagonal gluon fields but with an opposite sign compared with their counterparts in Eq.\eqref{34}.
Back to Eq.\eqref{33}, the last term can be rewritten with the help of Eq.\eqref{27},
	\begin{equation}
	\vec{F}^{\text{singular}}_{\mu\nu}.\vec{T}\equiv \dfrac{i}{g}N\left[ {\partial}_\mu, {\partial}_\nu\right]N^{\dagger}
	= \dfrac{1}{g} \left[ \partial_\mu ,\partial_\nu\right] \varphi T^3.
	\label{36}
\end{equation}
$F^{\text{singular}}_{\mu\nu}=\vec{F}^{\text{singular}}_{\mu\nu}.\vec{T}$ in the above equation indicates the field strength of an anti-thin vortex 
carrying a magnetic flux equal to $\Phi^{\text{flux}}=+\dfrac{2\pi}{g}T^3$.

\paragraph*{\textbf{Step 2: \textquotedblleft Center projection\textquotedblright}}

Adding $F^{\text{linear}}_{\mu\nu}$, $F^{\text{bilinear}}_{\mu\nu}$, and  $F^{\text{singular}}_{\mu\nu}$ together, one 
finds out that the similar terms with opposite signs cancel each other in $F^{\text{bilinear}}_{\mu\nu}$ and $F^{\text{linear}}_{\mu\nu}$.
On the other hand, the anti-thin vortex field  strength tensor contribution represented by $F^{\text{singular}}_{\mu\nu}$ is canceled by the thin vortex 
field strength tensor contribution brought in the 
last term of $F^{\text{linear}}_{\mu\nu}$, and finally, one is left with a full QCD field strength tensor.

\begin{figure}[ht]
	\begin{center}
		\centering
		\includegraphics[height=4cm, width=8cm]{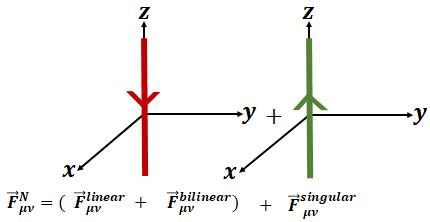}
		\caption{Appearance of a vortex and an anti-vortex under center gauge transformation.}
		\label{fig2}
	\end{center}
\end{figure}

In fact, with the above parametrization, one can argue that the vacuum is filled with thin vortices and anti-thin vortices, (see Fig.\eqref{fig2}).
	
To have only the contribution of the thin vortices, we remove the $F^{\text{singular}}_{\mu\nu}$ term in Fig.\eqref{fig2}. We call
this procedure \textquotedblleft center projection\textquotedblright\ and the \textquotedblleft center projected field strength tensor\textquotedblright\ 
is defined as follows:
\begin{widetext}\begin{equation}\begin{split}
			\vec{F}&^{\text{CP}}_{\mu\nu}.\vec{T}\equiv \vec{F}^{\text{linear}}_{\mu\nu}.\vec{T}+\vec{F}^{\text{bilinear}}_{\mu\nu}.\vec{T}
			\\
			&=\left[  \left( \partial_\mu A^1_\nu-\partial_\nu A^1_\mu\right) cos\varphi +\left( \partial_\mu A^2_\nu-\partial_\nu A^2_\mu\right) sin\varphi+g\left( A^1_\mu A^3_\nu-A^1_\nu A^3_\mu\right) sin\varphi-g\left( A^2_\mu A^3_\nu-A^2_\nu A^3_\mu\right) cos\varphi
			\right] T^1\\&+	\left[ - \left( \partial_\mu A^1_\nu-\partial_\nu A^1_\mu\right) sin\varphi +\left( \partial_\mu A^2_\nu-\partial_\nu A^2_\mu\right) cos\varphi+g\left( A^1_\mu A^3_\nu-A^1_\nu A^3_\mu\right) cos\varphi+g\left( A^2_\mu A^3_\nu-A^2_\nu A^3_\mu\right) sin\varphi
			\right]  T^2
			\\
			&+\left[  \partial_\mu A^3_\nu-\partial_\nu A^3_\mu+g\left( A^2_\mu A^1_\nu-A^1_\mu A^2_\nu\right) \right] T^3+\left( -\dfrac{1}{g}\left[ \partial_\mu, \partial_\nu\right] \varphi \right) T^3.
			\label{37}
\end{split}\end{equation}\end{widetext}
All the terms in Eq.\eqref{37} are regular except the last term, which represents the field strength of a thin vortex field.

Therefore, the gauge symmetry is SO($3$), and
thin vortices appear as the topological defects corresponding to the non-trivial first homotopy group $\Pi_1\left( \text{SO}(3)\right) =Z_2$.
\section{\label{sec:level5} indirect method of introducing vortices in the continuum}

In this section, motivated by the IMCG method in lattice QCD which confirms the existence of vortices, we study vortices in the continuum.
As mentioned in Sec. \ref{sub 2.2}, in the indirect method, in addition to the center gauge transformation and center projection \cite{DFG97}, an initial 
step including Abelian gauge transformation and then Abelian projection is done. 

Two successive gauge transformations are performed such that the first one is an Abelian gauge transformation $M(x)\in \text{SU}(N)$ and the second one is a center gauge 
transformation $N(x)\in \text{SU}(N)$. The transformation of link variables is
\begin{equation}
	U_\mu(x) \overset{M(x)}{\longrightarrow} U^M_\mu(x) \overset{N(x)}{\longrightarrow} U^{NM}_\mu(x)
	\label{40}
\end{equation}
or
\begin{equation}
	\begin{split}
		U^{NM}_\mu(x)&=N(x)U^M_\mu(x)N^{\dagger}(x+\hat{\mu})
		\\
		&=N(x)M(x)U_\mu(x) M^{\dagger}(x+\hat{\mu}) N^{\dagger}(x+\hat{\mu})
		\\
		&=N(x)M(x)e^{iagA_\mu} M^{\dagger}(x+\hat{\mu}) N^{\dagger}(x+\hat{\mu}).
		\label{41}
	\end{split}
\end{equation}
In the last equality, we used Eq.\eqref{4}.
Similar to what we have done in Sec. \ref{sub 3.1}, we can use the exponential expansion and 
by using the Taylor expansion for $M^{\dagger}(x+\hat{\mu})$ and $N^{\dagger}(x+\hat{\mu})$,

	\begin{widetext}
	\begin{equation}
		\begin{split}
			U^{NM}_{\mu}&=1+iag\left( N(x)\left[ M(x)A_\mu M^{\dagger}(x)-\dfrac{i}{g}M(x)\partial_\mu M^{\dagger}(x)\right]N^{\dagger}(x)-\dfrac{i}{g}N(x)\partial_\mu N^{\dagger}(x)\right) +\mathcal{O}(a^2)
			\\
			&=e^{iagA^{NM}_\mu}.
			\label{42}
		\end{split}
	\end{equation}
\end{widetext}
For the continuum limit where $a\rightarrow 0$,
\begin{widetext}
	\begin{equation}
		A^{NM}_\mu(x)=N(x)\left[M(x)A_\mu M^{\dagger}(x)-\dfrac{i}{g} M(x)\partial_\mu M^{\dagger}(x)\right]N^{\dagger}(x)-\dfrac{i}{g} N(x)\partial_\mu N^{\dagger}(x).
		\label{43}
	\end{equation}
\end{widetext}
$N(x)$ indicates a center gauge transformation, and the contribution of vortices must be obtained from this gauge transformation. 
Therefore, similar to the Sec. \ref{sub 3.1} and Eq.\eqref{15}, the gauge field  in the coset space is written as
\begin{widetext}\begin{equation}
		\vec{A}^{\prime NM}_\mu.\vec{T}=N(x)\left[M(x)\left( A^c_\mu T^c \right) M^{\dagger}(x)-\dfrac{i}{g} M(x)\partial_\mu M^{\dagger}(x)\right]N^{\dagger}(x)+\text{thin\ vortex}.
		\label{44}
\end{equation}\end{widetext}

This is somehow similar to an Abelian gauge fixing plus center gauge fixing of IMCG method in lattice calculations.
However, an intermediate step including Abelian projection must be applied, and we discuss  it for the SU($2$) gauge group in Sec. \ref{sec:level6}.

\section{\label{sec:level6} indirect method of introducing vortices in SU($2$) gauge group}

In this section, we apply the procedure explained in the previous section to the SU($2$) gauge group, and we show that, unlike Sec. \ref{sec:level4} where 
we have gotten vortices as a result of center gauge transformation and a \textquotedblleft center projection\textquotedblright, we get a chain of vortices and monopole by the indirect method.

We discuss this section in two subsections.
In Sec. \ref{sub 6.1}, we study the Abelian gauge transformation and Abelian projection, which leads to the emergence of the monopole.

In Sec. \ref{sub 6.2}, in addition to steps $1‌$ and $2$ in Sec. \ref{sub 6.1}, a center gauge transformation followed by a \textquotedblleft center projection\textquotedblright\
which leads to the appearance of chains containing vortices and monopoles, is discussed.

\subsection{Abelian gauge tTransformations and Abelian projection: Monopole}\label{sub 6.1}

The appearance of monopoles relies upon Abelian gauge transformation followed by an Abelian projection, which is discussed in this subsection for the SU($2$) gauge group.

Lattice studies show that within a good approximation the string tension between a pair of quark and anti-quark is described by Abelian variables of the 
maximal Abelian gauge transformation. Therefore, in the continuum limit, the idea of the Abelian gauge transformation is to repress the contribution 
of the off-diagonal components of the gauge fields so that the contribution of diagonal components is dominant in the low-energy regime.

We perform a local rotation in color space called an Abelian gauge transformation. As a result, magnetic monopoles can be extracted from the diagonal part of the non-Abelian gauge field.

\paragraph*{\textbf{Step 1: Abelian gauge transformation}}

Choosing $\alpha(x)=\varphi$, $\beta(x)=\theta$ and $\gamma(x)=\pm \varphi$ in the gauge rotation matrix of Eq.\eqref{26}, an appropriate 
gauge transformation $M(x)\in \text{SU}(2)$ which leads to an Abelian gauge transformation, is obtained.

In this paper, we choose $\gamma(x)=-\varphi$, 
\begin{equation}
	M(\theta,\varphi)=\left( \begin{matrix}
		cos\dfrac{\theta}{2}& e^{-i\varphi} sin\dfrac{\theta}{2}\\ -e^{i\varphi} sin\dfrac{\theta}{2} & cos\dfrac{\theta}{2}
	\end{matrix}\right).
	\label{45}
\end{equation} 
According to Sec. \ref{sub 3.1}, we define $G(x)\equiv M(x)$ as an Abelian gauge transformation; then the transformation of gluon field 
is given by Eq.\eqref{9},
\begin{equation}
	\vec{A}^M_\mu.\vec{T}=M(x)\left( \sum^3_{c=1}A^c_{\mu}T^c\right)  M^{\dagger}(x)-\dfrac{i}{g} M(x)\partial_\mu M^{\dagger}(x).
	\label{46}
\end{equation}
The first term on the right-hand side of Eq.\eqref{46} is regular under Abelian gauge transformation $M(x)$, but the second term is singular. 
Replacing Eq.\eqref{45} in Eq.\eqref{46},
\begin{widetext}\begin{equation}\begin{split}
			\vec{A}^M_\mu.\vec{T}&=\left[ A^1_\mu\left( 1-2sin^2\dfrac{\theta}{2}cos^2\varphi\right)+A^2_\mu\left(- sin^2\dfrac{\theta}{2}sin2\varphi\right)+A^3_\mu\left( -sin\theta cos\varphi\right)+\dfrac{1}{g}sin\varphi \partial_\mu\theta+\dfrac{1}{g}sin\theta cos\varphi \partial_\mu\varphi \right] T^1
			\\
			&+\left[ A^1_\mu\left( -sin^2\dfrac{\theta}{2}sin2\varphi\right)+A^2_\mu\left(1-2 sin^2\dfrac{\theta}{2}sin^2\varphi\right)+A^3_\mu\left( -sin\theta sin\varphi\right)-\dfrac{1}{g}cos\varphi \partial_\mu\theta+\dfrac{1}{g}sin\theta sin\varphi \partial_\mu\varphi \right] T^2
			\\
			&+\left[ A^1_\mu\left( sin\theta cos\varphi\right)+A^2_\mu\left( sin\theta sin\varphi\right)+A^3_\mu\left( cos\theta \right)+\dfrac{1}{g}\left( 1-cos\theta\right)\partial_\mu\varphi  \right] T^3,
			\label{47}
\end{split}	\end{equation}\end{widetext}
where the singularity appears in the inhomogeneous term of the above equation defined by $A^{\text{singular}}_\mu(\theta,\varphi)=A^{c\ \text{singular}}_\mu(\theta,\varphi)T^c\equiv-\dfrac{i}{g} M(\theta,\varphi)\partial_\mu M^{\dagger}(\theta,\varphi)$ in spherical coordinates and is given by
\begin{widetext}\begin{equation}\begin{split}
			\textbf{A}^{\text{singular}}(\theta,\varphi)&=-\dfrac{i}{g} M(\theta,\varphi)\nabla M^{\dagger}(\theta,\varphi)=\dfrac{1}{2g}\left( \begin{matrix}
				\left[ 1-cos\theta\right] \nabla\varphi&\left[  i\nabla\theta+sin\theta\nabla\varphi\right] e^{-i\varphi} \\\left[  -i\nabla\theta+sin\theta\nabla\varphi\right] e^{i\varphi} &- \left[  1-cos\theta\right] \nabla\varphi
			\end{matrix}\right)
			\\
			&=	\dfrac{g^{-1}}{r}\left(cos\varphi \textbf{e}_{\varphi}+sin\varphi \textbf{e}_{\theta} \right) T^1+\dfrac{g^{-1}}{r}\left(sin\varphi \textbf{e}_{\varphi}-cos\varphi \textbf{e}_{\theta} \right) T^2+\dfrac{g^{-1}}{r}\dfrac{1-cos\theta}{sin\theta}\textbf{e}_{\varphi}T^3,
			\label{48}
\end{split}\end{equation}\end{widetext}
where $\textbf{A}^{\text{singular}}(\theta,\varphi)=\textbf{A}^{c\ \text{singular}}(\theta,\varphi)T^c$.
It is observed from Eq.\eqref{48} that there exists a magnetic monopole as a point defect at the origin, $r=0$ along with a Dirac string at $\theta=\pi$.

The magnetic flux $\Phi^{\text{flux}}(\theta)$ of the inhomogeneous term is given by
\begin{widetext}\begin{equation}\begin{split}
			\Phi^{\text{flux}}(\theta)&=\int_c dx^{\mu}A^{singular}_\mu(\theta,\varphi) =\dfrac{1}{2g}\int^{2\pi}_0 d\varphi\left( \begin{matrix}
				1-cos\theta&\sin\theta e^{-i\varphi} \\sin\theta e^{i\varphi} &- \left(  1-cos\theta\right)  
			\end{matrix}\right)
			\\
			&=\dfrac{2\pi}{2g}\left( \begin{matrix}
				1-cos\theta&0\\0&- \left(  1-cos\theta\right)  
			\end{matrix}\right)=\dfrac{2\pi}{g} \left(  1-cos\theta\right)T^3.
			\label{49}
\end{split}	\end{equation}\end{widetext}
It is observed that the total contribution of the magnetic flux is located along the third direction of the color space. At $\theta=\pi$, 
the magnetic flux of a Dirac string that enters a monopole located at the origin $r=0$ is equal to $\dfrac{4\pi}{g}T^3$.

We have discussed in Sec. \ref{sub 3.2} that under a gauge transformation, the field strength tensor is obtained from Eq.\eqref{25}.
We rewrite it as follows:
\begin{equation}\begin{split}
		\vec{F}^M_{\mu\nu}.\vec{T}&= \left( \partial_\mu \left( \vec{A}^{ M}_\nu.\vec{T}\right) -\partial_\nu \left( \vec{A}^{M}_\mu.\vec{T}\right) \right)\\& +ig\left[\vec{A}^{ M}_\mu.\vec{T},\vec{A}^{M}_\nu.\vec{T} \right]\\&+\dfrac{i}{g}M(\theta,\varphi)\left[ {\partial}_\mu, {\partial}_\nu\right]M^{\dagger}(\theta,\varphi).
		\label{50}
\end{split}\end{equation}
Equation \eqref{50} can be calculated using Eqs.\eqref{45} and \eqref{47}.
Since the two color directions $T^1$ and $T^2$ have no contribution in the magnetic flux, we will suppress these non-diagonal components of the gauge fields in 
the \textquotedblleft Abelian projection\textquotedblright\ step.
It can be easily confirmed that the first term of the Abelian sector $\left( F^{\text{linear}}_{\mu\nu}\right)^3 \equiv \partial_\mu\left(A^M_\nu\right)^3-\partial_\nu\left( A^M_\mu\right)^3$ includes a magnetic monopole sitting at the origin along with a Dirac string in $\theta=\pi$.
The second term of the Abelian sector is called 
$\left(F^{\text{bilinear}}_{\mu\nu}\right)^3 \equiv -g\left\lbrace \left( A^M_\mu\right)^1\left( A^M_\nu\right)^2-\left(A^M_\mu\right)^2\left( A^M_\nu\right)^1 \right\rbrace $ and
contains an anti-monopole at the origin, and the third term of the Abelian sector 
$F^{\text{singular}}_{\mu\nu}\equiv\dfrac{i}{g}M(\theta,\varphi)\left[ {\partial}_\mu, {\partial}_\nu\right]M^{\dagger}(\theta,\varphi)$ includes an anti-Dirac string at $\theta=\pi$ with a magnetic flux equal to $-\dfrac{4\pi}{g}T^3$. (See Appendix A and Fig.\eqref{fig3}),

\begin{equation}
	 F^M_{\mu\nu}=F^{\text{linear}}_{\mu\nu}+F^{\text{singular}}_{\mu\nu}+F^{\text{bilinear}}_{\mu\nu}.
	\label{51}
\end{equation}

\begin{figure}[ht]
	\begin{center}
		\centering
		\includegraphics[height=3.75cm, width=8.68cm]{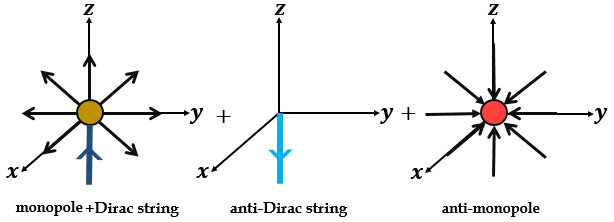}
		\caption{Appearance of a monopole accompanying a Dirac-string, an anti-Dirac string, and an anti-monopole as a result of an Abelian gauge transformation.}
		\label{fig3}
	\end{center}
\end{figure}  

If we add the contents of the above three terms,  the similar terms with opposite signs are canceled, and a field strength tensor which gives a full QCD is obtained \cite{Ichie}.

	\paragraph*{\textbf{Step 2: Abelian projection}}

The sum of the two terms $F^{\text{linear}}_{\mu\nu}+F^{\text{singular}}_{\mu\nu}$ represents a gauge configuration that only contains a monopole at $r=0$. However, it is exactly canceled by the anti-monopole arisen from $F^{\text{bilinear}}_{\mu\nu}$. Thus, one can claim that the vacuum is filled with monopoles and anti-monopoles.

From Fig.\eqref{fig3}, it is observed that, in order to have only the contribution of the monopole, we discard the term $F^{\text{bilinear}}_{\mu\nu}$ so 
that \cite{Ichie}
\begin{equation}
	\vec{A}^M_\mu.\vec{T}=\left( A^M_\mu\right)^aT^a \rightarrow \mathcal{A}_\mu\equiv \left( A^M_\mu \right) ^3 T^3.
	\label{52}
\end{equation}
From Eq.\eqref{47}, the gauge field is changed to
\begin{widetext}
	\begin{equation}
		\begin{split}
			\mathcal{A}_\mu= \left[ A^1_\mu\left( sin\theta cos\varphi\right)+A^2_\mu\left( sin\theta sin\varphi\right)+A^3_\mu\left( cos\theta \right)+\dfrac{1}{g}\left( 1-cos\theta\right)\partial_\mu\varphi  \right] T^3.
			\label{53}
		\end{split}
\end{equation}\end{widetext}
As a result, $F^{\text{bilinear}}_{\mu\nu}$,  which gives the anti-monopole contribution, is equal to zero, and the remaining part $F^{\text{linear}}_{\mu\nu}+F^{\text{singular}}_{\mu\nu}$ describes an Abelian projected QCD, which contains a monopole at $r=0$ and is called  the monopole vacuum.

After Abelian projection, SU($2$) gauge symmetry is reduced to U($1$) gauge symmetry, and
monopoles appear as the topological defects corresponding to the non-trivial second homotopy group $\Pi_2\left( \text{SU}(2)/\text{U}(1)\right) =\mathbb{Z}$.

\subsection{Center gauge transformation and \textquotedblleft Center Projection\textquotedblright: Chain}\label{sub 6.2}

As mentioned before, the vortex recognition by the indirect method is done in four steps. We have discussed  the first two steps in Sec. \ref{sub 6.1} 
and we explain the final two steps in the following.

\paragraph*{\textbf{Step 3: Center gauge transformation;}}

We have shown that under two successive gauge transformations the gluon field is changed by Eq.\eqref{44}.
After Abelian projection, the bracket in Eq.\eqref{44} should be replaced by Eq.\eqref{52},
\begin{equation}
	\vec{A}^{\prime NM}_\mu.\vec{T}=N(x) \mathcal{A}_\mu N^{\dagger}(x)+\text{thin\ vortex}.
	\label{54}
\end{equation}
For $x\notin$ hypersurface, the thin\ vortex is defined in Eq.\eqref{16}.

In fact, Eq.\eqref{54} expresses that a center gauge transformation is applied on a monopole vacuum.
Using the center gauge transformation defined by Eq.\eqref{27} and the Abelian projected field defined by Eq.\eqref{53}, the above equation is changed to
\begin{widetext}
	\begin{equation}\begin{split}
			\vec{A}^{\prime NM}_\mu.\vec{T}&=\left[ A^1_\mu\left(sin\theta cos\varphi \right)+ A^2_\mu\left(sin\theta sin\varphi \right)+A^3_\mu\left(cos\theta \right)+\dfrac{1}{g}\left( 1-cos\theta\right)  \partial_\mu\varphi-\dfrac{1}{g}\partial_\mu\varphi\right] T^3
			\\
			&=\left[ A^1_\mu\left(sin\theta cos\varphi \right)+ A^2_\mu\left(sin\theta sin\varphi \right)+A^3_\mu\left(cos\theta \right)-\dfrac{1}{g}cos\theta \partial_\mu\varphi\right] T^3.
			\label{55}
	\end{split}	\end{equation}
\end{widetext}
The first three terms on the right-hand side of Eq.\eqref{55} are regular, and the last term is defined in the spherical coordinates as
$\left( -\dfrac{1}{g}cos\theta \partial_\mu\varphi T^3=-\dfrac{g^{-1}}{r}\dfrac{cos\theta}{sin\theta} \textbf{e}_{\varphi}T^3\right)$. It indicates a defect representing a monopole located at the origin $r=0$ along with the two vortices at $\theta=0,\pi$.
This singular term remarkably represents a chain containing monopole and vortices.
In fact, the magnetic potential of the chain defined by $E_\mu\equiv -\dfrac{1}{g}cos\theta \partial_\mu\varphi T^3$ , can be interpreted as the sum of two terms: a magnetic potential of a monopole along with a Dirac string defined by $B_\mu\equiv \dfrac{1}{g}\left( 1-cos\theta\right) \partial_\mu\varphi T^3$  plus a magnetic potential of a 
vortex defined by $V_\mu\equiv -\dfrac{1}{g}\partial_\mu\varphi T^3$.

The magnetic flux $\Phi^{\text{flux}}(\theta)$ passing through a closed contour $C(r,\theta)$ is defined as 
\begin{widetext}\begin{equation}\begin{split}
			\Phi^{\text{flux}}(\theta)&=\int_c dx^{\mu}A^{\prime\ \text{singular}}_{\mu}=\dfrac{2\pi}{2g}\left( \begin{matrix}
				1-cos\theta&0\\0&- \left(  1-cos\theta\right)  
			\end{matrix}\right)-\dfrac{2\pi}{2g}\left( \begin{matrix}
				1&0\\0&- 1
			\end{matrix}\right)
			\\
			&=\dfrac{2\pi}{g} \left(  1-cos\theta\right)T^3-\dfrac{2\pi}{g}T^3
			\\
			&=-\dfrac{2\pi}{g}cos\theta T^3,
			\label{56}
\end{split}	\end{equation}\end{widetext}
where the first term in the second line of Eq.\eqref{56} represents the magnetic monopole flux plus a Dirac string. At $\theta=\pi$, the contribution of the flux of the Dirac string is equal to $+\dfrac{4\pi}{g}T^3$. The second term in the second line of Eq.\eqref{56} indicates a vortex extending on the $z$ axis with a flux equal to $-\dfrac{2\pi}{g}T^3$. Finally, the chain flux is obtained as the sum of the vortex flux and the
magnetic monopole flux plus the Dirac string and is equal to $-\dfrac{2\pi}{g}cos\theta T^3$.

Now, we have some discussions about the chain characteristic. From Eq.\eqref{56} for $\theta=0$, we only have the contribution of a magnetic vortex flux equal to $-\dfrac{2\pi}{g}T^3$, located in the positive direction of the $z$ axis which enters the magnetic monopole placed at the origin, $r=0$. 
At $\theta=\pi$, there exists a Dirac string flux equal to $+\dfrac{4\pi}{g}T^3$ located in the negative direction of the $z$ axis and entering the magnetic monopole. There is also a magnetic vortex whose flux is equal to $-\dfrac{2\pi}{g}T^3$ at $\theta=\pi$. It is located in the negative direction of the $z$ axis and exits from the magnetic monopole placed at $r=0$.
In fact, the sum of the two fluxes $\Phi^{\text{Dirac\ string}}+\Phi^{\text{line\ vortex}}$ represents the contribution of a vortex equal to $+\dfrac{2\pi}{g} T^3$, which enters the magnetic monopole sitting at the origin, $r=0$. 
As a result, the magnetic flux of the monopole is obtained as the sum of the absolute values of the fluxes of the two vortices entering into it.

Equation \eqref{25} is used to obtain the field strength tensor of the transformation,
\begin{widetext}
	\begin{equation}
		\begin{split}
			\vec{F}^{NM}_{\mu\nu}.\vec{T}&=N(x)\left( \vec{F}^M_{\mu\nu}.\vec{T}\right) N^{\dagger}(x)=N(x)M(x)\left( \vec{F}_{\mu\nu}.\vec{T}\right) M^{\dagger}(x)N^{\dagger}(x)
			\\
			&=\dfrac{1}{ig}\left[ \hat{D}^{NM}_\mu,\hat{D}^{NM}_\nu\right]-\dfrac{1}{ig}N(x)M(x)\left[ \hat{\partial}_\mu, \hat{\partial}_\nu\right] M^{\dagger}(x)N^{\dagger}(x)
			\\
			&= \left( \partial_\mu \left( \vec{A}^{\prime NM}_\nu.\vec{T}\right) -\partial_\nu \left( \vec{A}^{\prime NM}_\mu.\vec{T}\right) \right)  +ig\left[\vec{A}^{\prime NM}_\mu.\vec{T},\vec{A}^{\prime NM}_\nu .\vec{T}\right]+\dfrac{i}{g}N(x)M(x)\left[ {\partial}_\mu, {\partial}_\nu\right]M^{\dagger}(x)N^{\dagger}(x).
			\label{57}
		\end{split}
	\end{equation}
\end{widetext}
We use Eq.\eqref{57} to study some various topological defects.
A full QCD is obtained if one uses Eq.\eqref{57} without applying any projection.
In this section, we discuss the possible resulting defects after Abelian and \textquotedblleft center projections\textquotedblright. 

\begin{figure}[ht]
	\begin{center}
		\centering
		\includegraphics[height=4cm, width=8.30cm]{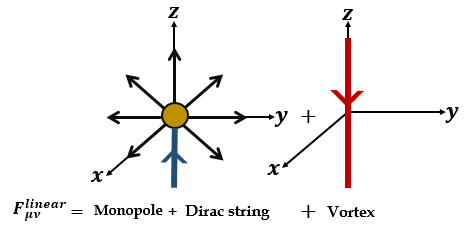}
		\caption{Appearance of a monopole attached to a Dirac-string and a vortex after Abelian gauge transformation, Abelian projection, and center gauge 
			transformation.}
		\label{fig4}
	\end{center}
\end{figure}

Looking at the last line of Eq.\eqref{57}, we define the first term by $F^{\text{linear}}_{\mu\nu}=\vec{F}^{\text{linear}}_{\mu\nu}.\vec{T}\equiv  \partial_\mu \left( \vec{A}^{\prime NM}_\nu.\vec{T}\right) -\partial_\nu \left( \vec{A}^{\prime NM}_\mu.\vec{T}\right)$ and rewrite it using Eq.\eqref{55},
\begin{widetext}\begin{equation}\begin{split}
			\vec{F}^{\text{linear}}_{\mu\nu}.\vec{T}&=\left\lbrace \left( \partial_\mu A^1_\nu-\partial_\nu A^1_\mu\right) sin\theta cos\varphi+A^1_\nu \partial_\mu\left(sin\theta cos\varphi \right)-A^1_\mu \partial_\nu\left(sin\theta cos\varphi \right)  \right\rbrace T^3
			\\
			&+\left\lbrace \left( \partial_\mu A^2_\nu-\partial_\nu A^2_\mu\right) sin\theta sin\varphi+A^2_\nu \partial_\mu\left(sin\theta sin\varphi \right)-A^2_\mu \partial_\nu\left(sin\theta sin\varphi \right)  \right\rbrace T^3
			\\
			&+\left\lbrace \left( \partial_\mu A^3_\nu-\partial_\nu A^3_\mu\right) cos\theta +A^3_\nu \partial_\mu\left(cos\theta \right)-A^3_\mu \partial_\nu\left(cos\theta \right)  \right\rbrace T^3
			\\
			&+\left( \partial_\mu E_\nu-\partial_\nu E_\mu\right)T^3.
			\label{58}
\end{split}\end{equation}\end{widetext}
The last line of the above equation contains some defects as explained in the following:
\begin{equation}\begin{split}
		\left( \partial_\mu E_\nu-\partial_\nu E_\mu\right)T^3&=\dfrac{1}{g}sin\theta\left(\partial_\mu\theta \partial_\nu\varphi-\partial_\mu\varphi\partial_\nu\theta \right)\ T^3\\& +\dfrac{1}{g}\left( 1-cos\theta\right) \left[ \partial_\mu,\partial_\nu\right] \varphi\ T^3\\&-\dfrac{1}{g} \left[ \partial_\mu,\partial_\nu\right] \varphi\ T^3.
		\label{59}
\end{split}\end{equation}
The first term of Eq.\eqref{59} represents the field strength of a magnetic monopole located at $r=0$, the second term indicates the field strength of a Dirac 
string at $\theta=\pi$, and the third term represents the field strength of a thin vortex field that extended on the $z$ axis. (See Fig.\eqref{fig4}.)

\begin{figure}[ht]
\begin{center}
	\centering
	\includegraphics[height=4cm, width=8.30cm]{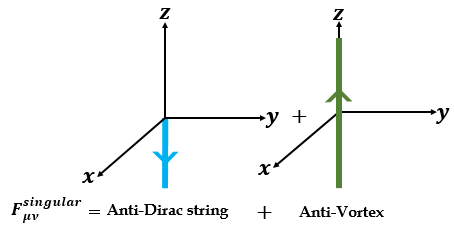}
	\caption{Appearance of an anti-Dirac string and an anti-vortex after Abelian gauge transformation, followed by an Abelian projection and then a center gauge transformation.}
	\label{fig5}
\end{center}
\end{figure}

It is clear that the second term of Eq.\eqref{57}, $\vec{F}^{\text{bilinear}}_{\mu\nu}.\vec{T}\equiv ig\left[ \vec{A}^{\prime NM}_\mu(x).\vec{T},\vec{A}^{\prime NM}_\nu(x).\vec{T}\right] $, is zero. 
This happens because of the Abelian projection process which makes the components of the gluon field zero for the two color directions $T^1$ and $T^2$. 
Using the center gauge transformation defined in Eq.\eqref{27} and the Abelian gauge transformation of Eq.\eqref{45}, the third term of Eq.\eqref{57} is
\begin{equation}\begin{split}
		F^{\text{singular}}_{\mu\nu}&=\vec{F}^{\text{singular}}_{\mu\nu}.\vec{T}\\&=-\dfrac{1}{g}\left( 1-cos\theta\right) \left[ \partial_\mu,\partial_\nu\right] \varphi\ T^3+\dfrac{1}{g} \left[ \partial_\mu,\partial_\nu\right] \varphi\ T^3\\&=\dfrac{1}{g} cos\theta\left[ \partial_\mu,\partial_\nu\right] \varphi\ T^3,
		\label{60}
\end{split}\end{equation}
where $-\dfrac{1}{g}\left( 1-cos\theta\right) \left[ \partial_\mu,\partial_\nu\right] \varphi\ T^3$ represents the field strength of an anti-Dirac string in $\theta=\pi$ with a flux equal to $-\dfrac{4\pi}{g}T^3$ and the term $\dfrac{1}{g} \left[ \partial_\mu,\partial_\nu\right] \varphi\ T^3$ represents the field strength of an anti-vortex on the $z$ axis 
with a flux equal to $+\dfrac{2\pi}{g}T^3$.  (See Fig.\eqref{fig5})

In fact, the contribution of the vortex and the Dirac string appearing in $F^{\text{linear}}_{\mu\nu}$ is exactly canceled by the contribution of the anti-vortex and the anti-Dirac string in $F^{\text{singular}}_{\mu\nu}$. As a result, a monopole vacuum is obtained unless we remove some of the singularities from the theory.

	\paragraph*{\textbf{Step 4: \textquotedblleft Center projection\textquotedblright}}

\begin{figure}[ht]
	\begin{center}
		\centering
		\includegraphics[height=3.75cm, width=8.52cm]{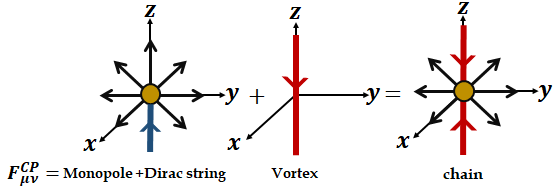}
		\caption{Appearance of a chain after Abelian gauge transformation, Abelian projection, center gauge transformation, and \textquotedblleft center projection\textquotedblright.}
		\label{fig6}
	\end{center}
\end{figure}

As explained in Sec. \ref{sec:level4}, a \textquotedblleft center projection\textquotedblright\ is done by removing $F^{\text{singular}}_{\mu\nu}$ defined in Eq.\eqref{60}. This means that \textquotedblleft center projection\textquotedblright\ is obtained by $F^{\text{linear}}_{\mu\nu}+F^{\text{bilinear}}_{\mu\nu}$. 
On the other hand, we have shown that $F^{\text{bilinear}}_{\mu\nu}$ is zero, and as a result,  the  center projected field strength tensor is as follows:
\begin{equation}
	F^{\text{CP}}_{\mu\nu}=F^{\text{linear}}_{\mu\nu}.
	\label{61}
\end{equation}
Therefore, only a monopole attached to a Dirac string and a vortex remain. We can interpret these configuration as a chain as shown in Fig.\eqref{fig6}.

The first plot on the left-hand side of Fig.\eqref{6} represents a monopole at $r=0$ plus a Dirac string located at $\theta=\pi$ carrying a magnetic flux equal to $\dfrac{4\pi}{g}T^3$. The second plot on the left-hand side indicates a vortex carrying a magnetic flux equal to $-\dfrac{2\pi}{g}T^3$ extending on the $z$ axis.
Combining these two plots, a chain shown on the right-hand side of Fig.\eqref{6} is obtained. A chain contains a monopole at $r=0$ and two vortices entering it. The flux of the vortex sitting at $\theta=0$ is equal to $-\dfrac{2\pi}{g}T^3$, and the flux of the vortex sitting at $\theta=\pi$ is equal to $+\dfrac{2\pi}{g}T^3$. The latter vortex is obtained as a result of combining the flux of the Dirac string sitting in the negative $z$ direction and the first vortex located in the $z$ direction. 

\begin{figure}[ht]
	\begin{center}
		\centering
		\subfloat[]{\includegraphics[height=1.5cm, width=8.5cm]{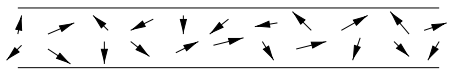}}
		\\
		\subfloat[]{\includegraphics[height=1.5cm, width=8.5cm]{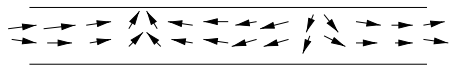}}
		\\
		\subfloat[]{\includegraphics[height=1.5cm, width=8.5cm]{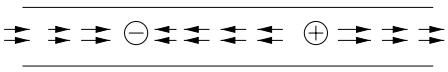}}
		\caption{Vortex field strength: (a) before gauge fixing, (b) after maximal Abelian gauge fixing in the horizontal $\pm \sigma^3$ direction, and 
                   (c) after Abelian projection \cite{DFGO}. As shown in this figure, two vortex lines enter a monopole or an anti-monopole, in agreement with 
                      what we have introduced in this paper as a chain configuration in Fig. \eqref{fig6}.}
		\label{fig7}
	\end{center}
\end{figure}

\begin{figure}[ht]
	\begin{center}
		\centering
		\subfloat[]{\includegraphics[height=2.5cm, width=3.5cm]{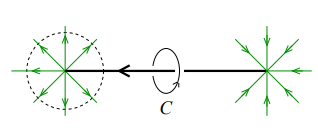}}
		\quad
		\subfloat[]{\includegraphics[height=2.5cm, width=3.5cm]{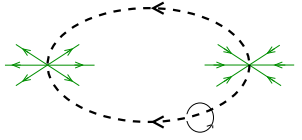}}
		\caption{ Illustration of the connection between Dirac string shown in plot (a) and the center vortex shown in plot (b) \cite{ER2000}.
                 The interpretation of a chain represented in Fig. \eqref{fig6} is the same as this figure where two vortex lines enter a monopole.}
		\label{fig8}
	\end{center}
\end{figure}

Our arguments about the chains of monopoles and vortices are in agreement with the work of Del Debbio \textit{et al}. \cite{DFGO}, which is done by lattice QCD (see Fig.\eqref{fig7}), 
and also in agreement with the results of Reinhardt and Engelhardt \cite{ER2000}, in which two vortex enter a monopole (see Fig.\eqref{fig8}).

We end this section by discussing the possible advantages of using chains.
As we mentioned at the beginning of the article, in both the dual superconductor model and the center vortex model, monopoles and vortices can explain some aspects of
the color confinement like the linear potential between a quark and anti-quark. 
However, none of these models nor the associated defects is able to describe all the 
expected features of the confining potential between color sources. 

At intermediate distances, a well-defined linear confining potential is expected, $V_R(r)\sim\sigma_R r$, in which $\sigma_R$ is the string tension of 
representation $R$.
The confining potential should agree with the Casimir scaling at intermediate distances.
It means that the string tension of the potential between a quark and an anti-quark in representation $R$, $\sigma_R$, is approximately proportional 
to the quadratic Casimir operator $C_R$ of representation $R$, $\text{i.e.}$,
\begin{equation}
	\sigma_R=\dfrac{C_R}{C_F}\sigma_F,
	\label{62}
\end{equation}
where $F$ indicates the fundamental representation and $\sigma_F$ shows the string tension of the fundamental representation.
$C_F$ denotes the eigenvalue of the Casimir operator of representation $F$.
We recall that the dependence of the potential slope to the Casimir scaling applies only for the intermediate distances and it is valid and exact for the large $N$ limit \cite{GH, Witten}.
In addition, at large distances, the $k$-string tension depends on the $N$-ality of the representation; it is equal to the fundamental representation string tension for the non-zero N-ality representations 
and zero for the zero $N$-ality representations \cite{kondo-n}.

Proportionality with Casimir scaling for the intermediate distances and the $N$-ality dependence of the potentials at large distances are confirmed by lattice calculations for the fundamental and a variety of higher representations \cite{deldar,deldar1,bali}. Therefore, any phenomenological model which tries to describe the potential between static color sources is expected to interpret these two features. Vortex based models have been able to explain the $N$-ality dependence. However, to get the Casimir scaling for all representations, the models have been modified by defining a thickness for the vortex \cite{DFGO, DFGO1}. On the other hand, lattice results confirm the existence of chains of monopoles and vortices \cite{AGG,monte} that
may explain the agreement of the potentials with Casimir scaling for higher representations. In this article, we 
have followed this approach to study the existence of chains of monopoles and vortices for the continuum.

We recall that an Abelian-projected theory gives the $N$-ality dependence (after all, it can 
still contain vortices), but it does not give the Casimir scaling dependence at intermediate distances \cite{DFGO97, final}

In this paper, motivated by direct and indirect methods of identifying vortices in lattice QCD, we have shown the existence of chains of monopoles and vortices 
for the continuum.

\section{\label{sec:level7}Conclusions}
Motivated by lattice QCD, which discusses the vortex contribution in color confinement, we have tried to introduce vortices in the continuum.

In the absence of matter fields, we work in the quenched approximation where dynamical quarks are removed from the theory. Therefore, the theory includes only the gluon fields, in this limit.

In recent years, the identification of vortices in lattice QCD has seen significant progress.
Therefore, one expects to observe the same physics in the continuum limit when one uses the lattice results for the limit where $a\rightarrow 0$.

Inspired by direct and indirect maximal center gauge methods which have studied vortices in lattice calculations and by using connection formalism technique, we 
have tried to recognize the vortices in the continuum. We have introduced the thin vortices from the gluon fields via 
both direct and indirect methods for the SU($N$) gauge group. We also get some help from the techniques proposed by Engelhardt and Reinhardt.

For an example, from the direct method, we have shown that under center gauge transformation the QCD vacuum of the SU($2$) gauge group is filled with the vortices and 
anti-vortices. Then, applying a \textquotedblleft center projection\textquotedblright\ we reach a theory that contains the thin vortex.  The theory has an SO($3$) symmetry containing the vortex, which corresponds to the non-trivial first homotopy group of $\Pi_1\left( \text{SO}(3)\right)=Z_2$.

Then, using the indirect method, we have shown that under Abelian gauge transformation for the SU($2$) case the gauge theory would contain monopoles 
along with the Dirac strings and anti-Dirac strings as well as anti-monopoles. Then, applying Abelian projection and removing the anti-monopole 
contribution, we end up with a theory that includes only the monopoles. In other words, SU($2$) gauge symmetry is reduced to a U($1$) gauge 
symmetry, and monopoles appear as the topological defects corresponding to the non-trivial second homotopy group $\Pi_2\left( \text{SU}(2)/\text{U}(1)\right)=\mathbb{Z}$.

Next, we have done a center gauge transformation on the Abelian vacuum. As a result, we get the monopole along with the Dirac string, the vortex, and anti-Dirac 
string, and the anti-vortex. Eventually, by applying \textquotedblleft center projection\textquotedblright, we end up with a theory that contains chains including monopole and two vortices.
\section*{Appendix}\label{appendix}

\subsection{Transformation of field strength tensor under an Abelian gauge transformation}

In subsection \ref{sub 6.1}, we express that the field strength is changed by Eq.\eqref{50} when Abelian gauge transformation is applied. One can also show that,
\begin{equation}\tag{A.1}\begin{split}
			\vec{F}^M_{\mu\nu}.\vec{T}&=\vec{F}^{\text{linear}}_{\mu\nu}.\vec{T}+\vec{F}^{\text{bilinear}}_{\mu\nu}.\vec{T}+\vec{F}^{\text{singular}}_{\mu\nu}.\vec{T}
			\\&
			\equiv
			\left( \partial_\mu \left( \vec{A}^{ M}_\nu.\vec{T}\right) -\partial_\nu \left( \vec{A}^{M}_\mu.\vec{T}\right) \right)\\& +ig\left[\vec{A}^{ M}_\mu.\vec{T},\vec{A}^{M}_\nu.\vec{T} \right]\\&+\dfrac{i}{g}M(\theta,\varphi)\left[ {\partial}_\mu, {\partial}_\nu\right]M^{\dagger}(\theta,\varphi).
			\label{A.1}
\end{split}\end{equation}
Each term of Eq.\eqref{A.1} can be expressed in terms of the SU($2$) group generators as the following:
\begin{widetext}\begin{equation}\tag{A.2}\begin{split}
			&\vec{F}^{\text{linear}}_{\mu\nu}.\vec{T}=\left( F^{\text{linear}}_{\mu\nu}\right)^1T^1+\left( F^{\text{linear}}_{\mu\nu}\right)^2T^2+\left( F^{\text{linear}}_{\mu\nu}\right)^3T^3,
			\\
			&\vec{F}^{\text{bilinear}}_{\mu\nu}.\vec{T}=\left( F^{\text{bilinear}}_{\mu\nu}\right)^1T^1+\left( F^{\text{bilinear}}_{\mu\nu}\right)^2T^2+\left( F^{\text{bilinear}}_{\mu\nu}\right)^3T^3,
			\\
			&\vec{F}^{\text{singular}}_{\mu\nu}.\vec{T}=\left( F^{\text{singular}}_{\mu\nu}\right)^1T^1+\left( F^{\text{singular}}_{\mu\nu}\right)^2T^2+\left( F^{\text{singular}}_{\mu\nu}\right)^3T^3.
			\label{A.2}
\end{split}\end{equation}\end{widetext}
From Eq.\eqref{47}, $\vec{F}^{\text{linear}}_{\mu\nu}.\vec{T}$ is obtained for the Abelian sector
\begin{widetext}\begin{equation}\tag{A.3}\begin{split}
			\left( F^{\text{linear}}_{\mu\nu}\right)^3&=\left( \partial_\mu A^1_\nu-\partial_\nu A^1_\mu\right) sin\theta cos\varphi+A^1_\nu \partial_\mu\left(sin\theta cos\varphi \right)-A^1_\mu \partial_\nu\left(sin\theta cos\varphi \right) 
			\\
			&+ \left( \partial_\mu A^2_\nu-\partial_\nu A^2_\mu\right) sin\theta sin\varphi+A^2_\nu \partial_\mu\left(sin\theta sin\varphi \right)-A^2_\mu \partial_\nu\left(sin\theta sin\varphi \right)  
			\\
			&+ \left( \partial_\mu A^3_\nu-\partial_\nu A^3_\mu\right) cos\theta +A^3_\nu \partial_\mu\left(cos\theta \right)-A^3_\mu \partial_\nu\left(cos\theta \right)
			\\
			&+\dfrac{1}{g}sin\theta\left(\partial_\mu\theta \partial_\nu\varphi-\partial_\mu\varphi\partial_\nu\theta \right)+\dfrac{1}{g}\left( 1-cos\theta\right) \left[ \partial_\mu,\partial_\nu\right] \varphi,
			\label{A.3}
\end{split}\end{equation}\end{widetext}
where, $+\dfrac{1}{g}sin\theta\left(\partial_\mu\theta\partial_\nu\varphi-\partial_\mu\varphi\partial_\nu\theta \right)$ represents  the field strength of a magnetic monopole at $r=0$ and $\dfrac{1}{g}\left( 1-cos\theta\right) \left[ \partial_\mu,\partial_\nu\right] \varphi$ indicates  the field strength of a Dirac string at $\theta=\pi$. 

The Abelian sector $\vec{F}^{\text{bilinear}}_{\mu\nu}.\vec{T}$ is also obtained by Eq.\eqref{47},
\begin{widetext}\begin{equation}\tag{A.4}\begin{split}
			\left( F^{\text{bilinear}}_{\mu\nu}\right)^3=&-g\left( A^2_\mu A^3_\nu- A^3_\mu A^2_\nu\right) sin\theta cos\varphi-A^1_\nu \partial_\mu\left(sin\theta cos\varphi \right)+A^1_\mu \partial_\nu\left(sin\theta cos\varphi \right) 
			\\
			&-g\left( A^3_\mu A^1_\nu- A^1_\mu A^3_\nu\right) sin\theta sin\varphi-A^2_\nu \partial_\mu\left(sin\theta sin\varphi \right)+A^2_\mu \partial_\nu\left(sin\theta sin\varphi \right)  
			\\
			&-g\left( A^1_\mu A^2_\nu- A^2_\mu A^1_\nu\right) cos\theta -A^3_\nu \partial_\mu\left(cos\theta \right)+A^3_\mu \partial_\nu\left(cos\theta \right)
			\\
			&-\dfrac{1}{g}sin\theta\left(\partial_\mu\theta \partial_\nu\varphi-\partial_\mu\varphi\partial_\nu\theta \right),
			\label{A.4}
\end{split}\end{equation}\end{widetext}
where $-\dfrac{1}{g}sin\theta\left(\partial_\mu\theta\partial_\nu\varphi-\partial_\mu\varphi\partial_\nu\theta \right)$ represents  the field strength of an anti-monopole at $r=0$.

Finally, $\vec{F}^{\text{singular}}_{\mu\nu}.\vec{T}$ is obtained for the Abelian sector by Eq.\eqref{45},
\begin{equation}\tag{A.5}
	\left( F^{\text{singular}}_{\mu\nu}\right)^3=-\dfrac{1}{g}\left( 1-cos\theta\right) \left[ \partial_\mu,\partial_\nu\right] \varphi.
	\label{A.5}
\end{equation}
And the above singular term shows  the field strength of an anti-Dirac string at $\theta=\pi$.

%	\subsection{\label{sec:level2}Second-level heading: Formatting}

%	\subsubsection{Wide text (A level-3 head)}

%	\paragraph{Note (Fourth-level head is run in)}
	
%	\subsection{\label{sec:citeref}Citations and References}
	
%	\subsubsection{Citations}
	
%	\paragraph{Syntax}

%	\paragraph{Eliding repeated information}

%	\paragraph{The options of the cite command itself}

\end{document}